%% file: kdd_2025_mtmd.tex
\documentclass[sigconf]{acmart}
\setcopyright{cc}
\setcctype[4.0]{by}  

\usepackage{svg}
\usepackage{multirow}
\usepackage{pifont}

\AtBeginDocument{%
  }

\setcopyright{acmlicensed}
\copyrightyear{2025}
\acmYear{2025}
\acmDOI{XXXXXXX.XXXXXXX}

\acmConference[AdKDD 2025]{31st ACM SIGKDD Conference on Knowledge Discovery and Data Mining}{03-07 August 2025}{Toronto, Canada}
\acmISBN{978-1-4503-XXXX-X/18/06}

\def\({{\big (}}
\def\){\big )}

\makeatletter
\newcommand*{\bigcdot}{}
\DeclareRobustCommand*{\bigcdot}{%
  \mathbin{\mathpalette\bigcdot@{}}%
}
\newcommand*{\bigcdot@scalefactor}{.5}
\newcommand*{\bigcdot@widthfactor}{1.15}
\newcommand*{\bigcdot@}[2]{%
  \sbox0{$#1\vcenter{}$}
  \sbox2{$#1\cdot\m@th$}%
  \hbox to \bigcdot@widthfactor\wd2{%
    \hfil
    \raise\ht0\hbox{%
      \scalebox{\bigcdot@scalefactor}{%
        \lower\ht0\hbox{$#1\bullet\m@th$}%
      }%
    }%
    \hfil
  }%
}
\makeatother

\makeatletter

\def\env@sqcases{
  \let\@ifnextchar\new@ifnextchar
  \left\lbrack
  \def\arraystretch{1.2}%
  \array{@{}l@{\quad}l@{}}%
}

\makeatother

\begin{document}

\title{MTMD: A Multi-Task Multi-Domain Framework \\
for Unified Ad Lightweight Ranking at Pinterest}

\author{Xiao Yang}
\affiliation{%
  \institution{Pinterest Inc.}
  \streetaddress{651 Brannan St}
  \city{San Francisco}
  \state{California}
  \country{USA}
}
\email{xyang@pinterest.com}

\author{Peifeng Yin}
\affiliation{%
  \institution{Pinterest Inc.}
  \streetaddress{651 Brannan St}
  \city{San Francisco}
  \state{California}
  \country{USA}
}
\email{pyin@pinterest.com}

\author{Abe Engle}
\affiliation{%
  \institution{Pinterest Inc.}
  \streetaddress{651 Brannan St}
  \city{San Francisco}
  \state{California}
  \country{USA}
}
\email{aengle@pinterest.com}

\author{Jinfeng Zhuang}
\affiliation{%
  \institution{Pinterest Inc.}
  \streetaddress{651 Brannan St}
  \city{San Francisco}
  \state{California}
  \country{USA}
}
\email{jzhuang@pinterest.com}

\author{Ling Leng}
\affiliation{%
  \institution{Pinterest Inc.}
  \streetaddress{651 Brannan St}
  \city{San Francisco}
  \state{California}
  \country{USA}
}
\email{lleng@pinterest.com}

\renewcommand{\shortauthors}{Yang et al.}

\begin{abstract}
The lightweight ad ranking layer, living after the retrieval stage and before the fine ranker, plays a critical role in the success of a cascaded ad recommendation system. Due to the fact that there are multiple optimization tasks depending on the ad domain, e.g., Click Through Rate (CTR) for click ads and Conversion Rate (CVR) for conversion ads, as well as multiple surfaces where an ad is served (home feed, search, or related item recommendation) with diverse ad products (shopping or standard ad); it is an essentially challenging problem in industry on how to do joint holistic optimization in the lightweight ranker, such that the overall platform's value, advertiser's value, and user's value are maximized. 

Deep Neural Network (DNN)-based multitask learning (MTL) can handle multiple goals naturally, with each prediction head mapping to a particular optimization goal. However, in practice, it is unclear how to unify data from different surfaces and ad products into a single model. It is critical to learn domain-specialized knowledge and explicitly transfer knowledge between domains to make MTL effective. We present a \textbf{M}ulti-\textbf{T}ask \textbf{M}ulti-\textbf{D}omain (MTMD) architecture under the classic Two-Tower paradigm, with the following key contributions: 1) handle different prediction tasks, ad products, and ad serving surfaces in a unified framework; 2) propose a novel mixture-of-expert architecture to learn both specialized knowledge each domain and common knowledge shared between domains; 3) propose a domain adaption module to encourage knowledge transfer between experts; 4) constrain the modeling of different prediction tasks. MTMD improves the offline loss value by 12\% to 36\%, mapping to 2\% online reduction in cost per click. We have deployed this single MTMD framework into production for Pinterest ad recommendation replacing 9 production models.

\end{abstract}



\begin{teaserfigure}
  \includegraphics[width=\textwidth]{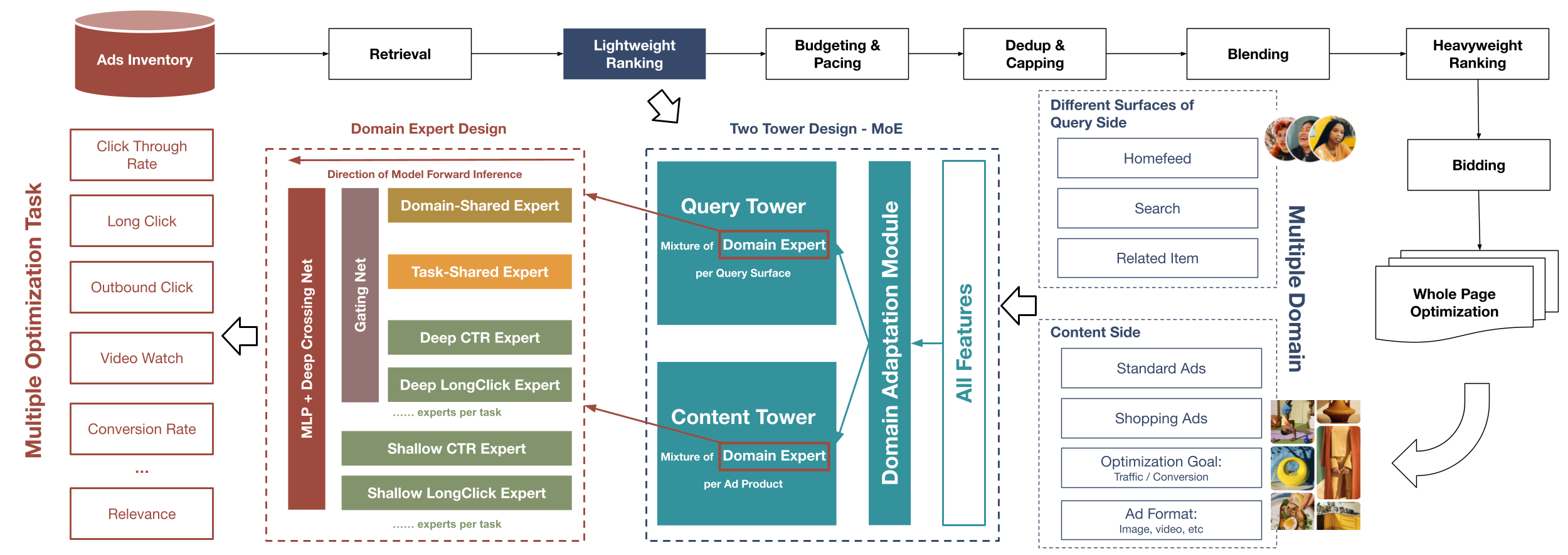}
  \caption{Illustration of online ad serving system, where the lightweight ranking is a middle tier optimizing all downstream goals. MTMD is a two-tower architecture that unifies both the input side (multiple domains) and the output side (multiple optimization tasks) based on a Mixture-of-Expert architecture.}
  \label{fig:teaser}
\end{teaserfigure}


\maketitle

\input{paper}

\bibliographystyle{ACM-Reference-Format}
\bibliography{kdd_2025_mtmd}

\end{document}

%% file: paper.tex
\section{Introduction}\label{sec:introduction}

The goal of modern online ad recommendation is to serve ad content that is highly personalized to user's interests, tastes, goals, and intent at the right time through the right funnel, such that user's value, advertiser's value, and platform's value are jointly maximized. 

\begin{figure}
\centering
\includegraphics[width=\columnwidth]{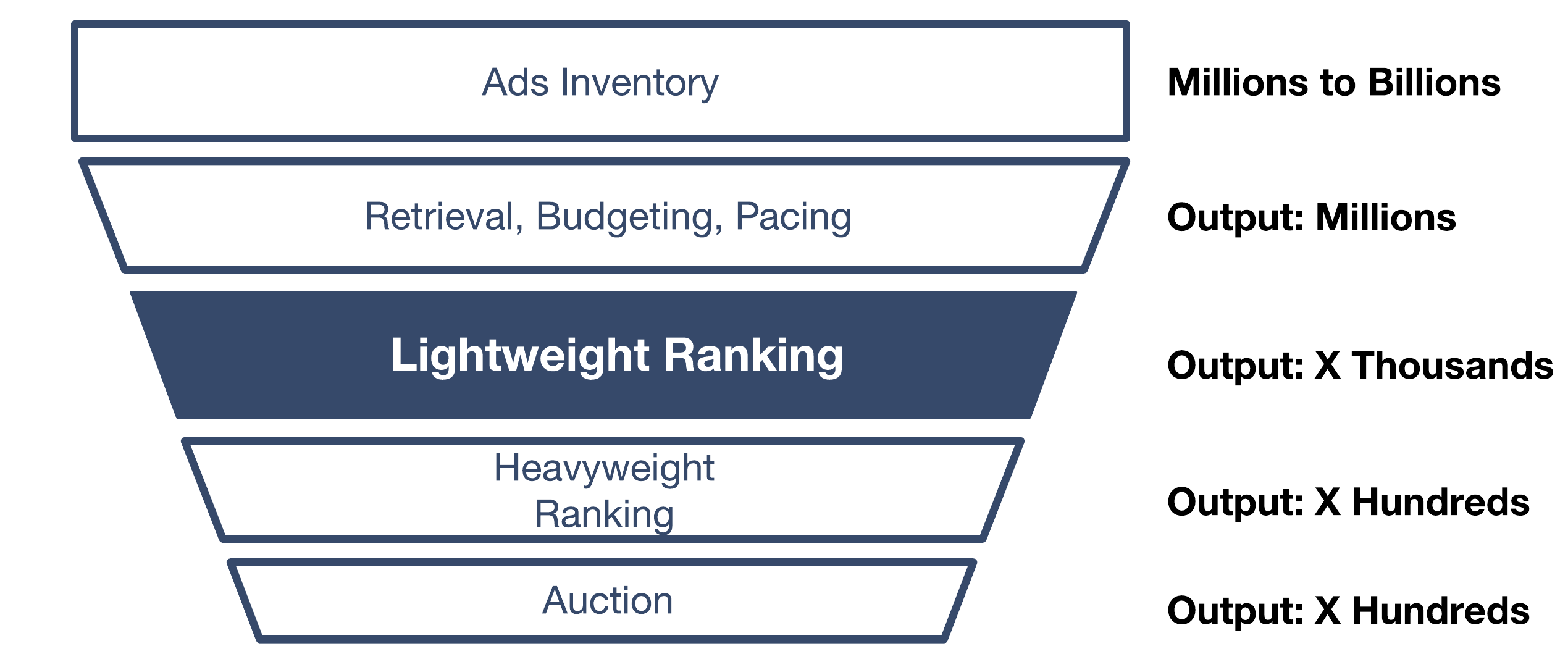}
\vspace{-0.5cm}
\caption{The throughput of each stage in a typical ad delivery funnel. Two-tower based model has the advantage of fast inference when the throughput is high.}
\label{fig:funnel}
\vspace{-0.5cm}
\end{figure}

Modern advertising recommendation is often a cascading system~\cite{GallagherCBC19}, divided into retrieval, lightweight ranking~\footnote{Also referred by "pre-ranking", "early ranking", "coarse ranking", "retrieval ranking" etc. We choose the name "lightweight" in this paper to emphasize the fact that latency plays an important role in our design choices.}, heavyweight ranking, and auction stages. The latency requirement of each layer is different depending on the number of ads entering and surviving that stage. Figure~\ref{fig:funnel} presents the scale in a typical delivery funnel. The lightweight ranking tier usually has millions to billions of ads as input. Due to the latency requirement, the lightweight ranker often uses a two-tower architecture~\cite{BromleyBBGLMSS93,HuangHGDAH13}, which allows the dot product between the query embedding and the item embedding as a fast inference function. Its performance is on the critical path to high-quality ad recommendation.

However, it is a very challenging problem to improve lightweight ranking in practice. The first challenge is the alignment of the delivery funnel: even if lightweight ranking is measured better in some scientific metrics, the gain often diminishes if the latter heavyweight ranking is not respecting the changes. Lightweight and heavyweight models are often not trained together. Second, there are many ad domains in a real-world system, e.g., the query side has different surfaces (home feed or search) and users, while the content side has different ad products, formats, and optimization goals. The lightweight ranker needs to handle them together as a single tier, with a model complexity budget to meet the latency requirement.

For the first alignment challenge, consistency-oriented lightweight ranking approaches have recently been proposed, to make ranking results similar to heavyweight ranking~\cite{QinZ0LLTZY022,TangW18,haoGZHLSWZJXZ23,WangJHZLZCZYWCC23}. Our MTMD learning uses the prediction score from the heavyweight ranker as the label, which encourages funnel alignment in nature. We limit the scope of this paper to the second multi-domain challenge.


One common practice is to train a model per ad domain. This approach suffers from two problems: 1) data fragmentation: a single training example was used in separate models, either for different tasks (e.g., CTR and CVR) or for different surfaces (e.g., home feed and search). Each domain could have more training data if we can share all training data; 2) maintenance and development velocity: we have to try all domains if we hope to test a new innovation or migrate some models. Furthermore, there are often mix shift phenomena between different models, where one task or one surface is improved while another is harmed.


Multitask learning (MTL)~\cite{MaZYCHC18,ZhaoHWCNAKSYC19,TangLZG20,WangJHZLZCZYWCC23} has been widely adopted to handle various optimization tasks jointly in industrial recommendation systems. However, naive handling of domain-specific features is suboptimal: When a feature is only available for a specific domain, the values of it are often set to certain default ones. Intuitively, having a more sophisticated domain adaptation module to handle missing values could improve model performance. When the input domains are very different from each other, both domain-specialized knowledge and common knowledge are required to maximize the power of MTL.

Specific to the model architecture, a general direction to improve two-tower models is to allow interaction modules between the query side and the item side~\cite{abs-2007-16122,Yu2021ADA,DongNBAW0Z22,SuYZXDZDT23,Li0GLZLL0GMLDT22,ZhaiGWSYLL23}. For our scale, we can only afford the dot product as the inference function due to latency consideration, and we need embedding output as a potential feature feeding to the heavyweight ranker. We do not explore these interaction modules for the focus of this paper.

Putting together, we propose MTMD, a \textbf{M}ulti-\textbf{T}ask \textbf{M}ulti-\textbf{D}omain network for unified web-scale lightweight ad ranker following the two-tower framework. The core building block of our network is a \textbf{Domain Expert}, which itself is a mixture of experts (MoE) to handle the challenges in a vanilla data-mixing MTL approach. The major value in the proposed method includes the following:
\begin{itemize}
    \item It has a series of fine-grained deep and shallow expert pairs for each prediction task, which explores the specialized knowledge of each task;
    \item It has domain-shared expert aiming to learn the common knowledge of different surfaces and ad products and allows more sophisticated domain adaptation to handle missing feature values;
    \item It has task-shared expert aiming to learn the common structure of different downstream prediction tasks;
    \item It has an expert routing layer to encourage different importance of experts.
\end{itemize}
With this Domain Expert as the foundational building block, the Query Tower in MTMD consists of $N$ Domain Experts mapping to each ad serving surface (e.g., home feed, search, etc.), and the Item Tower consists of $M$ Domain Experts mapping to each ad product type (e.g., shopping ad, traffic ad, etc.), followed by feedforward layers projecting tower output to the embedding vector for each task.
In addition, we introduce the constraint of different prediction tasks; for instance, a Good Click event (GCTR, click duration is longer than 30 seconds) depends on a Click event (CTR). 

We are able to deploy a single model in production replacing 9 lightweight ranking models on all surfaces, with significant gains in online metrics, including CTR ($\mathbf{+2.41\%}$), GCTR ($\mathbf{+3.06\%}$), CPC ($\mathbf{-1.96\%}$), and click volume ($\mathbf{+2.31\%}$), without increasing the impression of ads.

\section{Design of MTMD}\label{sec:design}

In this section, we present the network details of MTMD and the constrained modeling in MTL training.

\subsection{Domain Expert}\label{sec:de}

The design of Domain Expert focuses on the goal of unifying diverse input domains and output tasks, yet the resultant architecture should be composable and reusable for both the query side and the item side.

\subsubsection{Formulation of Ad Domain} 

The definition of "domain" is platform dependent. Specific to Pinterest ad recommendation, an ad domain means the diverse input training data and the output prediction tasks of the lightweight ranking tier. Specifically, it is keyed by three dimensions: 
\begin{itemize}
    \item the \textbf{Surface} that an ad is served, including \textit{home feed}, \textit{search}, and \textit{related item} recommendation;
    \item the \textbf{Ad Product} itself, including \textit{standard ad}, which aims for personalized events like impression, click, etc., and \textit{shopping ad}, which aims for offline conversion rate, i.e., a user purchases the item in this shopping ads;
    \item the \textbf{Prediction Task} of lightweight ranking, which includes \textit{click} (CTR), \textit{good/long click} (GCTR), \textit{outbound click} (OBC), \textit{conversion} (CVR), relevance, etc;
\end{itemize}

The total number of ad domain will be $|\text{Surface}| * |\text{Ad Product}|$, where $|\cdot|$ is the number of values of a concrete variable. This definition is general enough to be applied in other social media or e-cormerce platforms. MTMD aims to effectively handle this modeling diversity and complexity in a unified way so that all downstream heavyweight rankers can benefit from it. It should also significantly simplify the maintenance cost, compared to one model per domain. 

\subsubsection{Domain Adaptation} \label{sec:da}

\begin{figure}
\centering
\includegraphics[width=\columnwidth]{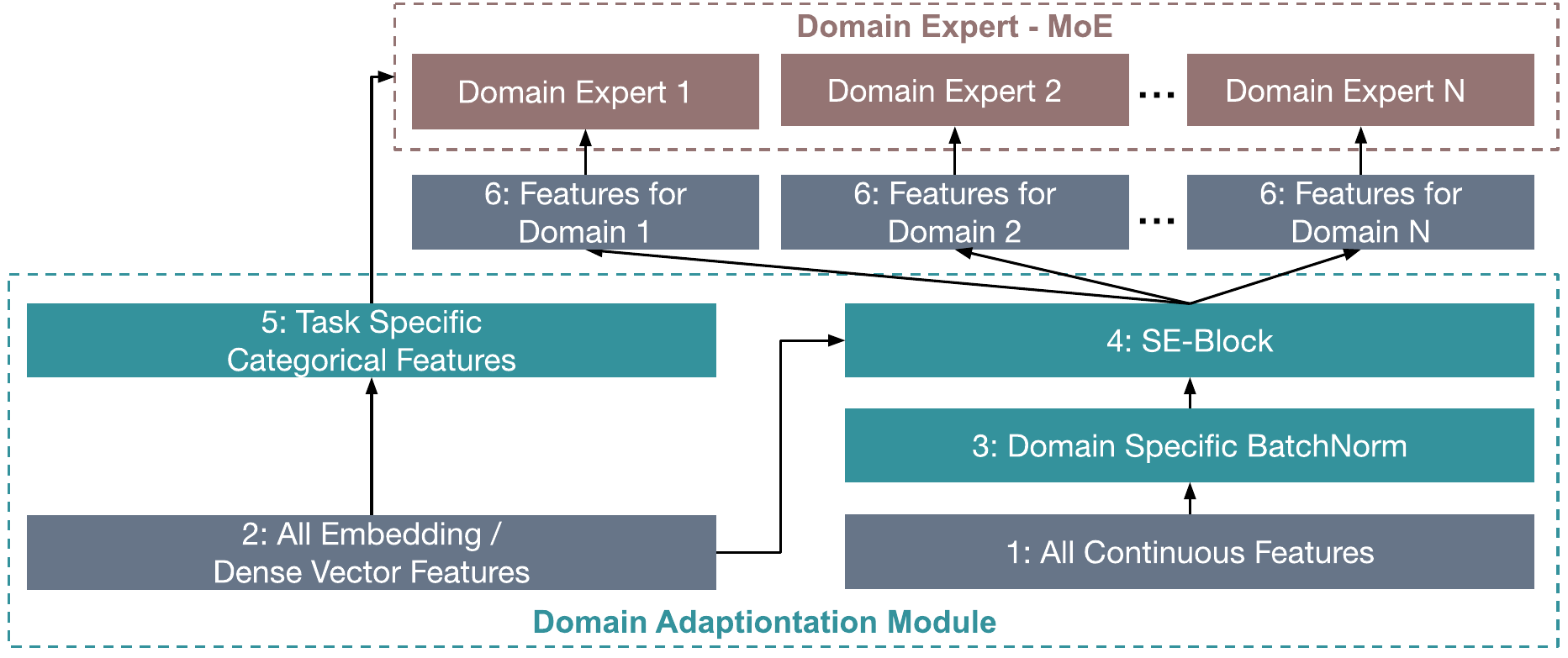}
\vspace{-0.3cm}
\caption{The design of the ad Domain Adaptation module in MTMD based on the Squeeze-and-Excitation block.}
\label{fig:adaptation}
\vspace{-0.5cm}
\end{figure}

The first challenge we face is how to process the features from different domains effectively. Our recipe is presented in the blocks of figure~\ref{fig:adaptation}. We use SE-Block from~\cite{JiangLZYLXDZ22} so that different tasks / domains can capture the difference in the importance of features in different domains. For all the continuous features, e.g. the click count of an ad in a past rolling window, we add one BATCH-NORM~\cite{IoffeS15} layer per domain before passing to the SE-Block.


\subsubsection{Task-Specific Experts}\label{sec:se}
Each task has a pair of deep and shallow experts:

\textbf{Deep Expert} consisting of a four-layer feedforward network (FFN) with dimensions $\{512, 256, 128, 128\}$. Each FFN layer consists of three sequential sublayers: fully connected linear projection~\cite{rosenblatt1958perceptron}, LAYER-NORM~\cite{ba2016layernormalization}, and LEAKY-RELU activation~\cite{xu2015empiricalevaluationrectifiedactivations} with negative slope $0.2$. The goal of this expert is to learn a deep domain-specific knowledge.

\textbf{Shallow Expert} consisting of a two-layer FFN of size $\{128, 64\}$. While the deep expert works on all input features after domain adaption, this shallow expert only takes some high-level categorical features that describe the key ad product properties, for example, the ad product type (traffic ads or conversion ads), etc.

The total number of task-specific experts is $2 * |\text{Prediction Task}|$.

\subsubsection{Domain-Shared and Task-Shared Expert}

We have both domain-shared expert and task-shared expert, each consists of a four-layer FFN of size $\{512, 256, 128, 128\}$, whose output will pass through an expert routing layer to help learning global common knowledge of different domains. This shared expert treatment also encourages task-specific experts to learn specialized knowledge within each domain.

\subsubsection{The Final Composition}\label{sec:comp}

With the three types of expert, that is, task-specific deep expert, task-shared expert and domain-shared expert, we use a routing layer to learn the importance of experts, followed by a DCN module to explore the power of feature crossing. Specifically, the final composition includes:

\textbf{Expert Routing} layer consisting of a three-layer FFN of size $\{128, 64, |\text{Expert}|\}$ followed by a \text{SOFTMAX} function. The purpose of this layer is to capture the importance of the output of each expert, assuming that different experts should contribute differently with their specialized knowledge. This expert routing idea has recently shown great success in large models~\cite{DaiDZXGCLZYWXLH24};

\textbf{Deep Crossing Network} (DCN) for feature crossing. It first applies LAYER-NORM to composited expert outputs, then applies the low-rank DCN module~\cite{WangSCJLHC21} on top of it to fully capture the crossing power of the feature crossing.

\textbf{Final Embedding Generation}: a linear layer that projects the output of DCN to the embedding $emb_{deep}$ per prediction task. Simultaneously, the output from the task-specific shallow expert, $emb_{shallow}$, is directly concatenated with $emb_{deep}$. Note that the dot product of the final embedding can be interpreted as the sum of the dot product of $emb_{deep}$ and the dot product of $emb_{shallow}$. This design allows the shallow expert to function as a calibration model, adjusting the logits from the deep networks. Finally, the output of Domain Expert is a dictionary where key is a prediction task and value is an embedding vector in Euclidian space. 

\begin{figure*}[t]
\centering
\includegraphics[width=\textwidth]{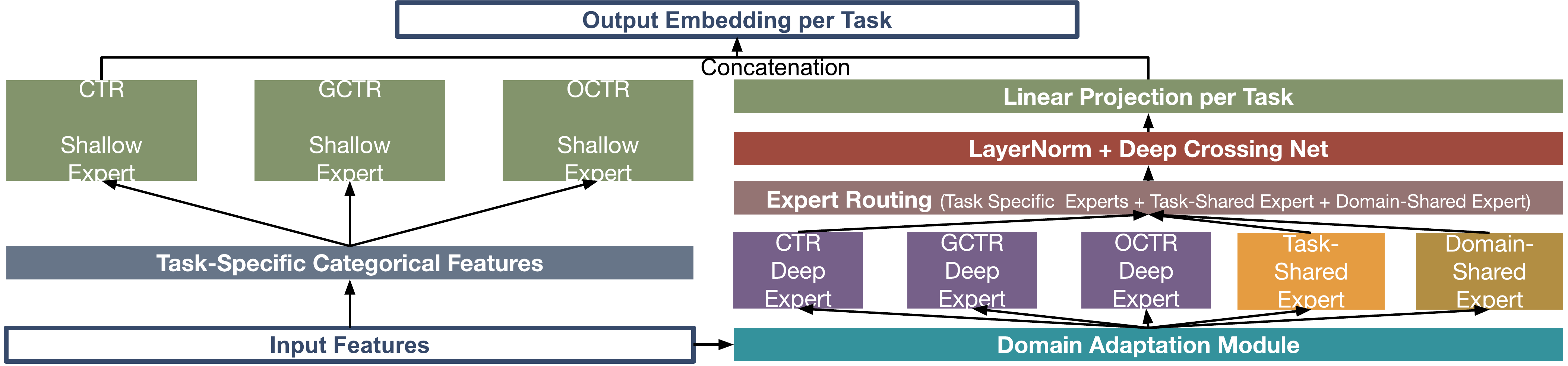}
\vspace{-0.5cm}
\caption{Design of Domain Expert. \textbf{On the Left:} shallow expert per task. \textbf{On the Right:} After passing Domain Adaptation processing of features, it goes through task specific Deep Expert, Task-Shared Expert, and Domain-Shared Expert.}
\label{fig:domain_expert}
\end{figure*}

Figure~\ref{fig:domain_expert} presents the hierarchy of expert composition with three tasks CTR, GCTR, and OCTR as examples. It can be generalized to other tasks including CVR and relevance, etc.

\vspace{-0.3cm}
\subsection{Query Tower and Item Tower}

With the Domain Expert design above, we are equipped to build our Query Tower and Item Tower following a two-tower architecture. 

\begin{figure}[t!]
\centering
\includegraphics[width=\columnwidth]{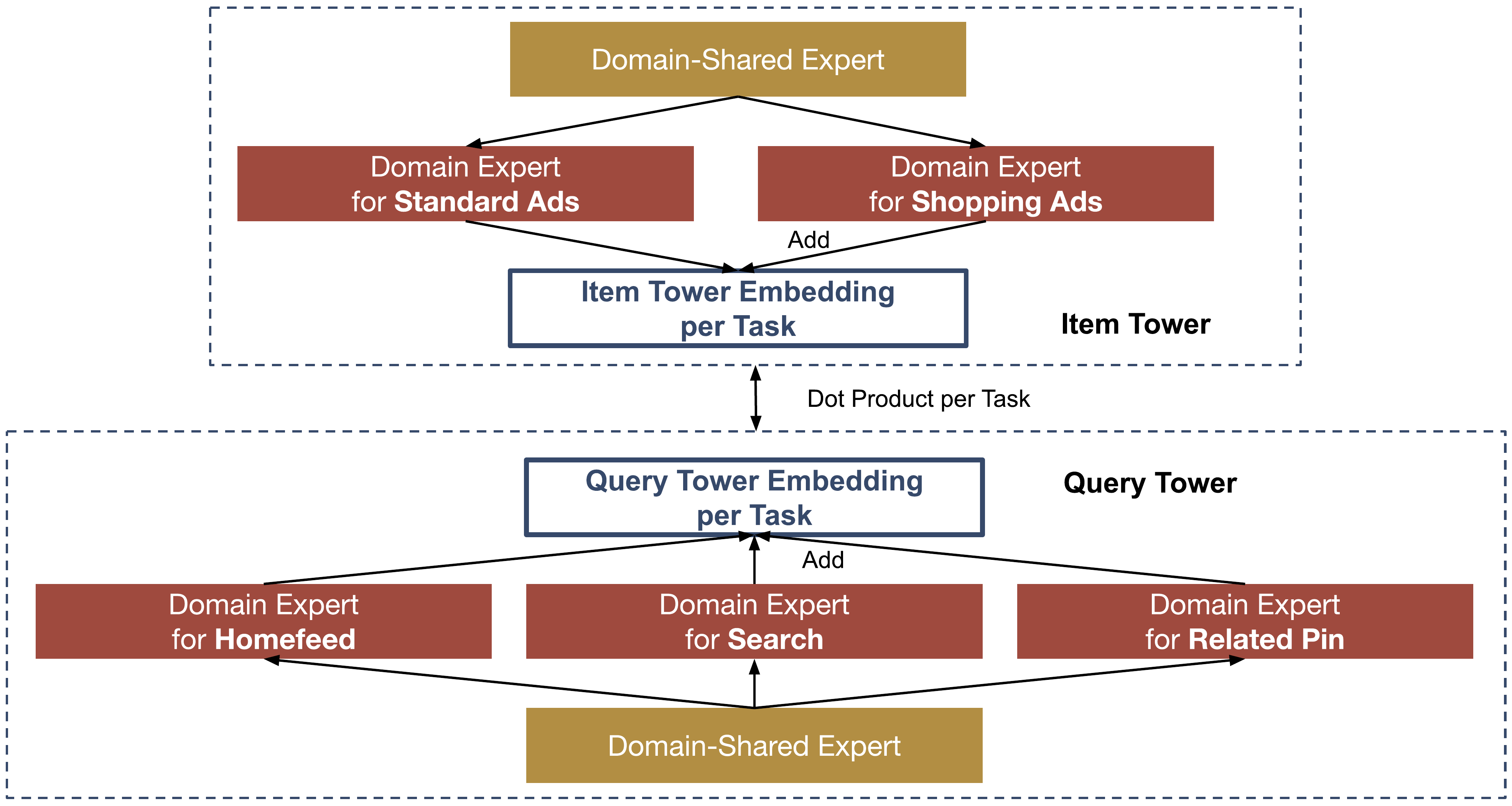}
\vspace{-0.5cm}
\caption{The two tower arch based on Domain Expert.}
\label{fig:two_tower}
\vspace{-0.5cm}
\end{figure}

Our key recipe is to decouple their types of domain experts from each other and design the tower architecture based on the uniqueness of Pinterest data. However, the idea is generalizable to other ad platforms, too: 
\begin{itemize}
    \item On the query side, we aggregate the Domain Expert \textbf{by different ad serving surfaces}: Home feed, Search, and Related Pins (related item recommendation in general);
    \item  On the item side, we aggregate the Domain Expert \textbf{by two different types of ad product}: Standard and Shopping ads. 
\end{itemize}
Each domain expert has its own collection of input features decided by the type of this domain expert. For example, a shopping ad expert on the item side will only receive the shopping features after the domain adaptation module. During serving time, only a single domain expert is activated for each request, which leads to improved infrastructure efficiency. Figure~\ref{fig:two_tower} summarized the two-tower architecture.

\subsection{Constrained Modeling}\label{sec:cm}

In a standard MTL (Multi-Task Learning) framework, each prediction task is equipped with its own Feed-Forward Network (FFN) layer to create embeddings specific to that task. These embeddings are then utilized in the dot product during inference, which essentially presumes that the tasks operate independently from one another. However, in our scenario, such a configuration ignores the conditional relationship between different prediction tasks. Taking the three types of clicks in Figure~\ref{fig:domain_expert} as an example, a long click (GCTR) should be conditioned on a click (CTR). The same is true for outbound clicks (OCTR). Moreover, an equal size of the output embedding dimension, which is 64 each task as in section~\ref{sec:de}, does not reflect the importance of the more fundamental role of CTR.

We propose a Constrained Modeling between different tasks to solve the above two limitations. First, we introduce the dependency between tasks in the formulation:

\begin{align*} 
\vspace{-0.3cm}
P(\text{CTR}) &= P(\text{Click}|\text{Impression}), \\ 
P(\text{GCTR}) &= P\big(\text{GCTR}|\text{Impression}\big) \Rightarrow P\big(\text{GCTR}|\text{CTR}\big) * P(\text{CTR}), \\
P(\text{OCTR}) &= P\big(\text{OCTR}|\text{Impression}\big) \Rightarrow P\big(\text{OCTR}|\text{CTR}\big) * P(\text{CTR}),
\end{align*}

In implementation, we use the KL-Divergence~\cite{10.1214/aoms/1177729694} between lightweight prediction score and the heavyweight prediction score as a loss function. By formulating GCTR and OCTR conditionally on CTR, the model can develop better understanding of CTR. GCTR and OCTR can also get causal benefits. Moreover, we used 128 dimensions for the CTR embedding, 32 for both the GCTR and the OCTR embedding, instead of an even 64 dimension for each task. We also use a larger weight on the loss of CTR prediction to further increase its importance in training.

\section{Experiments and Results}\label{sec:exp}

\subsection{Dataset and Experiment Setup}

Our offline experiments are conducted on a subset of a large-scale production dataset. This evaluation data set has approximately hundreds of millions of samples and tens of millions of unique users per day. The training periods for our models range from 20 to 60 days, ensuring sufficient data for robust model training and evaluation. The online experiments are based on the online A/B experiment to evaluate the top-line business metrics.


The ground truth label in training is the heavyweight ranker prediction score instead of the user action labels. MTMD can be applied to both engagement models and conversion models. It has been launched for both traffic and conversion ads at Pinterest. We focus on the three engagement tasks in this section. 

\vspace{-3mm}
\subsection{Offline Evaluation}

We evaluated on three surfaces: home feed (HF), search (SR), and related Pin (RP), 2 ad product type: standard ads and shopping ads, and 3 engagement tasks: CTR, GCTR, and OCTR. The baseline are production two-tower models trained using data only on a single slice. Each tower in baseline is a multilayer feedforward network after feature pre-processing. 

We explored KL-Divergence, LogMAE, MAE, BCE, etc. as offline evaluation metrics. Empirically, we have found that logMAE is stable and maps well to the online metric movement directionally. When it is positive in offline evaluation tables, it means it reduces LogMAE more.

\begin{table}
 \center
  \caption{Offline improvement of LogMAE of MTMD on different ad domains for three engagement tasks. A positive number means it reduces LogMAE further.}
  \vspace{-0.3cm}
  \begin{tabular}{rcccc}
    \toprule
    Surface	&Ad Type	&CTR	&GCTR	&OCTR \\
    \toprule
    home feed &shopping	&35.65\%	&32.30\%	&N/A\\
    home feed &standard	&20.25\%	&36.66\%	&23.40\%\\
    \midrule
    related Pin &shopping	&32.31\%	&31.06\%	&N/A\\
    related Pin &standard	&23.38\%	&22.65\%	&36.58\%\\
    \midrule
    search &shopping	&16.84\%	&23.48\%	&N/A\\    
    search &standard	&12.42\%	&23.69\%	&35.62\%\\    
    \bottomrule
  \end{tabular}
  \vspace{-0.5cm}
\label{table:offline}
\end{table}

%

Table~\ref{table:offline} presents the change in the offline metric. We can observe that MTMD significantly improved performance in all tasks and all domains by a large margin between 12\% and 36\%. The MTMD model possesses a comparable number of embedding parameters and a reduced number of non-embedding parameters relative to all baseline models combined. This robust offline performance provides compelling evidence that a singular, unified model can surpass individual models by integrating fragmented data and employing a meticulously designed model architecture.

These offline results are very encouraging. To further assess the model performance in real-world large-scale scenario, we conduct an online A/B experiment to understand the impact on online business metrics such as CTR and CPC.

\subsection{Online A/B Test}

\begin{table}[t!]
 \center
  \caption{Home feed VS Related Pin VS Search - Online A/B test results by ad serving surface.}
  \vspace{-0.3cm}
  \small
  \begin{tabular}{rcccccccccc}
    \toprule
    	&CPC &CTR &GCTR &OCTR &HDR &RPR\\
    \toprule
    Home feed	&\textbf{-2.95\%} &\textbf{3.62\%}	&3.29\% &4.77\% &-0.21\% &7.73\%\\
    \midrule
    Related Pin  &\textbf{-1.60\%} &\textbf{2.07\%}	&3.07\% &2.32\% &-0.13\% &3.27\%\\
    \midrule
    Search  &\textbf{-1.60\%} &\textbf{2.07\%}	&3.07\% &2.32\% &-0.13\% &3.27\%\\
    \bottomrule
  \end{tabular}
\label{table:ad_surface}
\vspace{-3mm}
\end{table}

\begin{table}[t!]
 \center
  \caption{Standard ads VS Shopping ads - Online A/B test results by ad product type.}
  \vspace{-0.3cm}
  \small
  \begin{tabular}{rcccccccccc}
    \toprule
    	&CPC &CTR &GCTR &OCTR &HDR &RPR\\
    \toprule
    Standard	&\textbf{-1.79\%} &\textbf{2.17\%}	&2.84\% &2.72\% &-0.01\% &4.46\%\\
    \midrule
    Shopping	&\textbf{-2.12\%} &\textbf{2.99\%}	&3.73\% &2.93\% &-2.45\% &5.19\%\\
    \bottomrule
  \end{tabular}
\label{table:ad_type}
\vspace{-3mm}
\end{table}

\begin{table}[t!]
 \center
  \caption{Click ads VS Conversion ads - Online A/B test results by ad optimization goal.}
  \vspace{-0.3cm}
  \small
  \begin{tabular}{rcccccccccc}
    \toprule
    	&CPC &CTR &GCTR &OCTR &HDR &RPR\\
    \toprule
    Click	&\textbf{-1.97\%} &\textbf{3.49\%}	&4.67\% &4.30\% &0.12\% &7.78\%\\
    \midrule
    Conversion	&\textbf{-1.88\%} &\textbf{1.87\%}	&2.18\%	&1.79\% &-0.71\% &3.07\%\\
    \bottomrule
  \end{tabular}
\label{table:ad_op}
\vspace{-3mm}
\end{table}

\begin{table}[t!]
 \center
  \caption{Constrained Modeling - Online A/B test results in percentage with MTMD model architecture as baseline.}
  \vspace{-0.3cm}
  \small
  \begin{tabular}{rcccccccccc}
    \toprule
    	&CPC &CTR &GCTR &OCTR &HDR &RPR\\
    \toprule
     Home feed	&\textbf{-0.81}\% &\textbf{0.61}\%	&0.79\% &0.62\% &-2.81\% &0.65\%\\
     Related Pin	&0.02\% &\textbf{0.43\%}	&-0.18\% &0.45\% &-6.15\% &0.22\%\\
     Search	&0.37\% &-0.18\%	&-0.65\% &-0.17\% &4.37\% &-1.03\%\\
    \midrule
     Standard	&-0.08\% &0.15\%	&-0.23\% &0.13\% &-3.86\% &-1.13\%\\
     Shopping	&\textbf{-0.37\%} &\textbf{0.60\%}	&0.56\% &0.63\% &-1.86\% &0.87\%\\
    \bottomrule
  \end{tabular}
\label{table:cm}
\vspace{-3mm}
\end{table}

We evaluate the following business metrics in an online A/B setup, where users are randomly split into control and treatment groups. The control group is a production model and the treatment group is MTMD.
\begin{itemize}
    \item CPC: cost per click on ad. Measures how much an advertiser pays for each click on their ad. The lower the better;
    \item CTR: ad clickthrough rate. It is important business metric for user experience;
    \item GCTR: the rate an ad is clicked and a user stays >30 seconds on it;
    \item OCTR: CTR for the ad that goes to a 3rd party site;
    \item HDR: hide rate of this ad. The lower the better;
    \item RPR: ad repin rate. It is a Pinterest-specific feature that strongly correlates with user experience;
\end{itemize}
Note that we do not track revenue, presumably the most important ad metric, because we force identical budget over control and treatment groups to eliminate the impact of budget cannibalization. We highlight the metric changes in CPC and CTR in bold font because they are the most crucial, indicating the business value and user value, respectively.

\textit{Significant Platform-level Lift} Overall MTMD reduces CPC and significantly increases CTR, which is one of the most significant launches in years for the lightweight ranking tier. The three prediction tasks CTR, GCTR, and OCTR are all improved at the same time, while the repin rate is significantly improved by $3+\%$ and the hide rate is slightly decreased. The online gain is also consistent across

\begin{itemize}
    \item Serving Surface: all three surfaces saw a significant CPC reduction with an increase in CTR;
    \item Product Type: table~\ref{table:ad_type} showed that CPC and CTR are improved significantly on both standard and shopping ad;
    \item Optimization Goal: MTMD increases the CTR of the click ad by $\mathbf{+3. 49\%}$, and reduces the CPC of the conversion ads by $\mathbf{-1.88\%}$. 
\end{itemize}

\textit{Performance by Constrained Modeling} The offline gain of constrained modeling is marginal after the 5th bucket. We conduct a separate online A/B experiment with MTMD as the baseline, to determine whether constrained modeling can further enhance the value of MTMD. Table~\ref{table:cm} shows that constrained modeling can bring statistically significant gains in CPC and CTR.

\vspace{-0.3cm}
\subsection{Ablation Study}
We present ablation studies measured by LogMAE to justify design choices. One factor is removed at a time to assess how the offline metric would change. The results are summarized in Table~\ref{table:ablation}. We did not conduct online A/B experiments for this purpose because of the large number of variants.
  \vspace{-0.4cm}
\begin{table}[h!]
\small
 \center
  \caption{Offline improvement of LogMAE of differet factors on the top ranked buckets.}
  \vspace{-0.4cm}
  \begin{tabular}{rcccccc}
    \toprule
    &0  &1  &2  &3  &4  &all\\
    \toprule
    DomainAdapt	&3.71\%	&3.53\%	&3.67\%	&3.88\%	&4.38\%    &\textbf{3.86\%}\\
    DCN V2	&0.90\%	&0.99\%	&1.15\%	&1.34\%	&1.71\%  &\textbf{1.24\%}\\
    Normalization	&2.36\%	&2.50\%	&2.64\%	&2.64\%	&2.84\%   &\textbf{2.61\%}\\
    DownSampling	&-0.23\%   &-0.29\%	&-0.84\%     &0.15\%	&-0.12\%  &-0.26\%\\
    \bottomrule
  \end{tabular}
\label{table:ablation}
\end{table}

The order of impact is Domain Adaptation > Proper Normalization > DCN Feature Crossing. The smaller embedding dimension and the downsampling of the data can slightly hurt the performance. Specifically,

\begin{itemize}
    \item Domain adaptation: brings $\mathbf{+3.86\%}$ improvement to LogMAE. This confirms the motivation that the transfer of feature representations between domains helps MTL;
    \item DCN crossing: the DCN V2 crossing module applied to the output of deep experts further improves the loss by $\mathbf{+1.24\%}$;
    \item Proper Normalization: contributes to $\mathbf{+2.61\%}$ loss improvement, Pre-norm is better than post-norm;
    \item Larger embedding dimension is better: We evaluated different dimensions of the embeddings of the two-tower output. It shows a clear trend for performance-wise $64>48>32$.
\end{itemize}

\section{Conclusion and Future Works}\label{sec:conclusion}

In this paper, we present a unified Multi-Task Multi-Domain (MTMD) lightweight ranking framework. The core design component is a Domain Expert, which adapts feature processing with respect to various domains, ensembles four different types of expert, and crosses the expert outputs using DCN V2. With this Domain Expert, we composite the pin tower by the type of ad product and the query tower by the surface an ad is served. This MTMD model replaced nine models in production and delivered a significant reduction in CPC and a substantial increase in CTR during the online A / B experiment. MTMD has been the lightweight ranking model in production at Pinterest.

In the future, we will improve our design in the two orthogonal directions: The first is the \textbf{alignment of the delivery funnel}, where the goal is to make the output of lightweight ranking and heavyweight ranking more consistent for better ad delivery efficiency. The other direction is to \textbf{allow better interaction} between the query tower and the pin tower to make the model more effective in capturing the personalized needs of the users. 

\section{Acknowledgments}

We thank Hari Venkatesan for his support in the development of MTMD.
